\UseRawInputEncoding
\documentclass[
 reprint,
superscriptaddress,
 amsmath,amssymb,
 aps,
pra,
]{revtex4-1}

\usepackage{graphicx}% Include figure files
\usepackage{dcolumn}% Align table columns on decimal point
\usepackage{bm}% bold math
\usepackage[caption=false]{subfig}
\usepackage{epsfig}

\begin{document}

%\preprint{APS/123-QED}

\title{Electronic correlations effect on nontrivial topological fermions in CoSi
}% Force line breaks with \\

\author{Paromita Dutta}
 \altaffiliation{dutta.paromita1@gmail.com}%Lines break automatically or can be forced with \\
 \affiliation{%
School of Basic Sciences, Indian Institute of Technology Mandi, Kamand, Himachal Pradesh-175075, India}%

\author{Sudhir K. Pandey}
\altaffiliation{sudhir@iitmandi.ac.in} 
\affiliation{%
School of Engineering, Indian Institute of Technology Mandi, Kamand, Himachal Pradesh-175075, India
}%

\date{\today}% It is always \today, today,

\begin{abstract}

The present study has been carried out to understand the effect of electronic correlations on the recently found fermions in CoSi. For which the spectral functions of bulk and (001) surface of CoSi have been investigated by using DFT+DMFT advanced methodology at T = 100 K with and without inclusion of spin-orbit coupling (SOC). All the newly found fermions are observed at $\Gamma$ and $R$ points similar to other theoretical and experimental reports. Our DFT+DMFT calculations for the bulk states have shown one extra hole pocket at $M$ point. Both coherent and incoherent features have been observed in the spectra of bulk CoSi. This indicates the presence of quasiparticle-quasiparticle (QP-QP) interactions which is eventually affecting the lifetime ($\tau$) of exotic fermionic QPs. For instance, the calculated $\tau$ for QPs at $\omega \sim $ -30 and -186 mev are found to be $\sim 10^{-9}$ s and $\sim 10^{-12}$ s, respectively when SOC is not considered. However, $G_0W_0$ corrections have shown $\tau$ for spin-1 fermionic QP at $\Gamma$ to be infinite while for double Weyl fermionic QP at $R$ point to be $\sim 10^{-12}$ s. Their effective masses ($m^*$) have also been calculated as $\sim$ 1.60 and 1.64 at $\Gamma$ and $R$ points, respectively. Furthermore, the spectral functions at T = 100 K of (001) surface have also shown both coherent and incoherent features. Consequently, at $\omega$ = 0 for surface states $\tau$ has been calculated of the order $\sim 10^{-8}$s for both without SOC and with SOC inclusions.

\end{abstract}

\maketitle

\section{Introduction}

For past few decades there has been a quest for novel fermionic particles other than Dirac, Majorana and Weyl in the field of elementary-particle physics \cite{Pal}. Interestingly, in the recent years condensed matter systems have also shown the existence of various exotic fermionic quasiparticles (QPs) such as two-dimensional (2D) Dirac fermions in graphene \cite{Castro}, helical Dirac fermions at the surface of three-dimensional (3D) topological insulators \cite{Hasan,Ando}, Majorana fermions in topological superconductors \cite{Sato,Zou,Li,Deng,Das}, Dirac semimetals and Weyl semimetals \cite{Wang,Weng,Neupane,Yang,Miao,-Y,Liu,Souma,Gibson,Mele}. Exploration of these kind of fermionic QPs is very important due to their potential of applicability in the field of electronics and information technology.

Following the history, Herring investigated the degeneracies occur in electronic band structures, and he noted that even in the absence of any symmetry one could obtain a two-fold degeneracy in a 3D solid \cite{Herring}. The dispersion in the vicinity of these degeneracy points (the band crossing/touching points) is generally linear and resembles the Weyl equation which lacks the Lorentz invariance \cite{Herring}. Such band touching points in solids as discovered by Herring are then named as Weyl nodes/points by Wan \textit{et al.} \cite{Herring,Wan}. The key feature of determining such band touching points is the degeneracy of bands which in turn is realized by symmetry. The Weyl points can be found in 3D solids only when either time reversal or inversion symmetries are broken. On other hand, when both the symmetries are present, the possibility of a 2-fold degeneracy at a Weyl point in the spectrum gets excluded. This rather produces a four-fold degeneracy at band crossing points which is also termed as Dirac node \cite{Mele}. Thus, at the band touching points Weyl fermions have 2-fold degeneracy and the Dirac fermions have 4-fold degeneracy. However, recent studies have shown new types of massless fermions with many-fold degeneracies in solid state systems which have no analogs in high energy physics \cite{Cano,Kim,Fang,Zhu,Dai,Lv}. These newly found fermions appear in crystals with specific space group symmetries \cite{Tang}. For example, spin-1 fermion with 3-fold degeneracy carrying $\pm$2 topological charge, spin-3/2 Rarita-Schwinger-Weyl fermion (RSW) with 4-fold degeneracy carrying $\pm$4 topological charge \cite{Ezawa,Liang,Tang}, double Weyl fermions with $\pm$2 topological charge \cite{Bouhon,Gilbert,Xu} and double spin-1 fermion with 6-fold degeneracy \cite{Cano}. Interestingly, these newly found fermions are predicted to be seen in topological chiral systems like transition-metal (TM) monosilicides such as XSi(X=Co,Rh,Fe) \cite{Tang,Gilbert,Shekhar,Sanchez,Takane,Rao,Bose}. Among these the present work is mainly focused onto CoSi for the study.
First-principle band structure calculations over CoSi and its angle-resolved photo-emission spectroscopic study (ARPES) have already predicted and shown it to be topological chiral systems \cite{Tang,Chang,Takane,Pshenay}; belongs to the family of CoGe, RhSi and RhGe with P2$_1$3 as spacegroup \cite{Flicker}. The 3D band structure calculations without spin-orbit coupling (SOC) for CoSi has predicted two types of unconventional linear band crossing points, one at $\Gamma$ with 3-fold degeneracy corresponding to spin-1 fermion and other at $R$ with 4-fold degeneracy corresponding to double Weyl fermion \cite{Tang,Pshenay}. Tang \textit{et al.} have also shown that topological surface states (SS) become apparent from the projections of spin-1 excitation and a double Weyl fermion at $\Gamma$ and $R$ \cite{Tang}. Moreover, when band structure calculations are performed with SOC consideration; resulted into two separate crossing points both at $\Gamma$ and $R$, respectively \cite{Tang}. Thus, recent works are mainly focused into the prediction and realizing the existence of unconventional chiral fermions and their associated fermi-arc states for CoSi material theoretically and experimentally \cite{Tang,Chang,Takane,Pshenay,Pan,Sakai,Ishii,Rao}.

All these theoretical works have been carried out at the density functional theory (DFT) level. In the non-interacting picture QPs possess infinite lifetime. Thus, in those DFT studies the newly found QPs have infinite lifetime, and the informations of their interactions are missed out. It is well known that in solids QP-QP interactions broaden the single noninteracting particle $\delta$-function peaks, renormalize them and redistribute the spectral weight between the coherent and incoherent structures. These incoherent structures often called statelites or sidebands. Thus, it is important to characterize these exotic fermionic QPs in interacting picture and which can be achieved by using beyond DFT method i.e., dynamical mean field theory (DMFT). For dealing many electron problems and strong correlation effects in solids DFT+DMFT is a sophisticated formulation \cite{Anisimov,Lichtenstein,Kotliar,Dang}, has been used recently \cite{Dutta_2019,Antik,Korotin,Kunes}. This formulation deals with local impurity problem, where self-energy ($\Sigma(\omega)$) contains all the information regarding QP excitations \cite{Imada}. However, these days $GW$ calculations are also performed for getting the QP's informations but this method is computationally very costly. This $GW$ approximation is a many-body perturbation theory as developed by Hedin \cite{Hedin}. Here, G stands for one electron Green's function and W stands for screened Coulomb interaction. In this theory, the self-energy ($\Sigma$) is a function of both crystal momentum ($\textbf{k}$) and frequency ($\omega$), and can calculate $\Sigma$ of QPs at any $k$-point in Brillouin zone (BZ). However, in this theory one-shot $GW$ ($G_0W_0$) is well established which is a all electron $GW$ based technique, and not much computationally demanding \cite{Pshenay,Antik}. Thus, usage of these two advanced methodologies will provide an insight of these new fermions' behavior inside strongly correlated electron system like CoSi. This kind of investigation is required because a bridge between the existing unconventional chiral fermions with nontrivial topology and the electronic correlations is missing. The present work is motivated with this thought, and accordingly, we have revisited the already found topological chiral fermions in CoSi with the usage of advanced formulations.

Considering the above mentioned aspects we have studied the spectral functions for both bulk and (001) surface of CoSi by using DFT+DMFT methodology at T = 100 K \cite{Haule} with and without SOC inclusions. All the new fermions are observed at $\Gamma$ and $R$ points as expected yet one extra hole pocket is found at $M$ points. DFT+DMFT calculations have shown the presence of both coherent and incoherent features in spectra. Suggesting the lifetime ($\tau$) of the newly found fermions to be affected due to QP-QP interactions. For instance, the calculated $\tau$ for QPs at $\omega \sim $ -30 and -186 mev are found to be $\sim 10^{-9}$ s and $\sim 10^{-12}$ s, respectively when SOC is not considered. However, $G_0W_0$ corrections have given $\tau$ for spin-1 fermionic QP at $\Gamma$ to be infinite while for double-weyl fermionic QP at $R$ point to be $\sim 10^{-12}$ s. Similarly, $\tau$ is calculated for other new fermionic QPs as found when SOC is considered by using both DFT+DMFT and $G_0W_0$ methods. Moreover, at T = 100 K the spectral functions of (001) surface have also shown the existence of both coherent and incoherent features. Accordingly, at $\omega$ = 0 $\tau$ has been evaluated which is of the same order $\sim 10^{-8}$s for both noc SOC and SOC considerations.

\section{Computational Details}

The DFT based electronic structure calculations have been performed by using WIEN2k code \citep{Blaha}. This code is based on full-potential plane wave (FP-LAPW) method. PBEsol is taken as exchange functional for the calculations \cite{Perdew}. The experimentally observed crystal structure is taken from literature \cite{B}. 2.18 and 1.84 Bohr are the muffin-tin sphere radii for Co and Si sites, respectively, with 10$^{-3}$ as charge convergence. 21$\times$21$\times$21 mesh grid size in BZ has been used for the bulk states calculations. Next, DFT+DMFT calculations have been performed by using the code as implemented by Haule \textit{et al.} \cite{Haule} which is interfaced with WIEN2k code \cite{Blaha}. This DMFT code provides stationary free energies at finite temperatures \cite{Birol}. Accordingly, DMFT calculations are carried out for 100 K temperature and all the calculations are performed fully self-consistently in the impurity levels and electronic charge density. The auxiliary impurity problem is solved by using a continuous-time quantum monte carlo impurity solver here \cite{K}. Exact double-counting scheme as proposed by Haule has been used here \cite{Haule_2015}. More informations regarding this DFT+DMFT code can be found at Ref.\cite{Haule,Kotliar,Pandey,dmft}. Full 3\textit{d} orbitals of Co are treated at DMFT level. The density-density form of Coulomb repulsion has been employed here with the usage of self-consistently calculated values of \textit{U}(4.5 eV) and \textit{J}(0.94 eV) from our previous work \cite{Dutta}. The analytical continuation as needed for obtaining the self-energy on the real axis maximum entropy method is used here \cite{Jarrell}. Further, to calculate the QP energies at particular $k$-points, 1 shot-GW ($G_{0}W_{0}$) calculations are performed by using the Questaal package \cite{Pashov}. This code is based on full-potential linearized muffin-tin orbital (FP-LMTO) method\cite{Methfessel,Kotani_2010}. In this code $G_{0}W_{0}$ calculations are perturbations to a DFT calculation. They are simpler than quaisparticle self consistent $GW$ calculations, because only the diagonal part of $\Sigma^0$​​ is normally calculated (this is an approximation) and only one self-energy is calculated (single iteration). Its $GW$ implementation description can be found in Ref.\cite{Kotani_2007}. Here again, the exchange functional and the muffin-tin sphere radii for Co and Si sites are kept same as mentioned above. 10$\times$10$\times$10 mesh grid for DFT calculations in BZ while 4$\times$4$\times$4 mesh grid for the self-energy in $G_{0}W_{0}$ calculations have been used. Lastly, the surface states of 001 surface are calculated at both DFT and DMFT levels. For this, a slab of thickness 33.54 bohr has been taken containing 32 atoms. To minimize the interactions between two consecutive slabs a 30 Bohr vacuum has been provided along $k_z$ direction. Moreover, SOC is considered in all the calculations.

\section{Results and Discussion}

Fig. 1 (a) illustrates the crystalline lattice structure of CoSi which crystallizes in a cubic lattice with non-symmorphic space group P2$_{1}$3 and lacks inversion symmetry. Here Co and Si atoms are located at 4a sites with position coordinates at (x,x,x) in a unit cell with x$_{Co}$=0.140 and x$_{Si}$=0.843. The corresponding Brillouin zone (BZ) along high-symmetric lines is shown in Fig. 1(b), which is a cube with the $\Gamma$-point at the center, R-points at the vertices, X-points at the centers of the faces and M-points at the centers of the edges of the cube. The projected BZ of 001 surface along high-symmetric lines ($\overline{M}$-$\overline{\Gamma}$-$\overline{X}$)is shown in Fig. 1(c).

Fig. 2(a) shows the bulk band structure as obtained from DFT, and Fig. 2(b) shows momentum-resolved many-body spectral function calculated within DFT+DMFT at temperature (T) = 100 K, along high-symmetric $k$-directions. Fig. 2(a) - 2(b) are plotted without SOC considerations in the energy window -0.5$\leq\omega\leq$0.2 eV. Bands are numbered in the Fig. 2(a) for the discussion purpose. Before stating the differences between the two plots, here it is important to understand that there is a difference in studying the electronic dispersion curves as generated from DFT and DFT+DMFT methods. Generally, the spectral function $(A(\textbf{k},\omega))$ is defined as $A(\textbf{k},\omega)=-Im G^R(\textbf{k},\omega)/\pi$ where $G^R(\textbf{k},\omega)$ is the retarded Green's function for the interacting electron system. $G^R(\textbf{k},\omega)= 1/(\omega-\epsilon_0(\textbf{k})- \Sigma(\textbf{k},\omega)$, where $\Sigma(\textbf{k},\omega)$ is the self-energy term in which all the interaction effects are contained. However, for the non-interacting system $\Sigma(\textbf{k},\omega)$ term is zero and $(A(\textbf{k},\omega))$ has a $\delta$-function peak at $\omega = \epsilon_0(\textbf{k})$. As DFT represents the non-interacting electron picture; here $A(\textbf{k},\omega)= \delta(\omega-\epsilon_0(\textbf{k}))$. On the other-hand within DFT+DMFT, the $A(\textbf{k},\omega)$ is written as given in Eq. 1 \cite{Imada}:
\begin{equation}
A(\textbf{k},\omega) = \dfrac{1}{\pi} \dfrac{-Im \Sigma(\textbf{k},\omega)}{[\omega - \epsilon_0(\textbf{k})- Re\Sigma(\textbf{k},\omega)]^2 + [-Im \Sigma(\textbf{k},\omega)]^2}
\end{equation}
where, $\omega$ is real frequency, $\epsilon_0(\textbf{k})$ is the single non-interacting electron's energy with crystal momentum ($\textbf{k}$), $Im \Sigma(\textbf{k},\omega)$ is the imaginary part and $Re \Sigma(k,\omega)$ real part of the self-energy, respectively. This $A(\textbf{k},\omega)$ will have one major peak with some finite width and rest spectrum will have broadened structures. The major peak is associated with DFT $\delta$ peak with broadened shape at an energy position ($\omega = \epsilon_0(\textbf{k})+Re \Sigma(\textbf{k},\omega)$); this peak will correspond to coherent weight. The rest broadened structures in the spectrum will be corresponded to incoherent weight.Now, when this spectrum is seen on a larger scale with large numbers of $n$ and k, the whole spectrum then have smeared attributions with sharp dispersive lines. The sharp dispersive lines are then associated with coherent weights (states with large lifetime) and smeared features of the spectrum with incoherent weights(states with shorter lifetime).

Here, it is important to note that in this study single site DFT+DMFT calculations have been performed, and thus the self-energy will have only $\omega$ dependence. $ Im \Sigma (\omega) $ contains the information regarding the lifetime of QPs. Larger is the value of $ Im \Sigma(\omega) $ lesser will be the lifetime of QPs. However, to get a better picture of the lifetime of QPs at particular $k$-point $G_0W_0$ calculations are performed in this study. Further with the knowledge of $ Re \Sigma(\omega) $, the effective band mass renormalization parameter ($m^* = 1 - d Re \Sigma / d\omega|_{\omega = 0}$) and QP weight ($Z=1/m^*$) can also be calculated. This $m^*$ provides the information of the renormalization of bands due to the inclusion of Coulomb interactions; resulting in a spectral weight transfer between the incoherent and coherent states. Generally, $Z$'s value remains $<$ 1 and positive for the interacting electron systems instead of 1 which is for the case of non-interacting system. Thus, more closer the value of $Z$ to 1 lesser will be the transfer of spectral weight from coherent states to incoherent states. Lastly, by following the Eq. 1, the lifetime ($\tau$) of QP can be calculated as $\tau \approx \hbar/$FWHM; where FWHM is full width half maximum \cite{Landau}, and in Eq. 1 FWHM is $\approx$ 2\,$Im \Sigma(\textbf{k},\omega)$.

Now, on the basis of above explanations, the two figures Fig. 2(a) \& 2(b) are compared. Firstly, well defined dispersive lines in Fig. 2(a) while in Fig. 2(b) smeared features (in the range -0.5 eV to -0.3 eV) with sharp dispersive lines (in the range -0.3 eV to 0.2 eV) are observed. Thus, in the range -0.2$\leq\omega\leq$0.2 eV QPs seem to be coherent (larger lifetime) and when $\omega \leq$ -0.2 eV the states are becoming incoherent. This can also be observed from Fig. 3(a), where the value of Im$\Sigma(\omega)$ is negligibly small in the range -0.2$\leq\omega\leq$0.2 eV for all the three 3\textit{d} components of Co. Secondly, in Fig. 2(a) one hole pocket around $\Gamma$ point (created by bands 1 \& 2) and one electron pocket around $R$ point (created by bands 3 \& 4) are observed. On the other hand in Fig. 2(b), two hole pockets are seen around $\Gamma$ and $M$ points (created by bands 1 \& 2), respectively, and one electron pocket at $R$ point (created by bands 3 \& 4). This appears that due to electronic correlations extra hole pockets seem to be generated at $M$ points. However, \textit{ab-initio} calculations as performed by Sanchez \textit{et al.} have reported hole pocket at $M$ point which is unlike Tang \textit{et al.} report \cite{Sanchez,Tang}. This suggests the urge of observing hole pocket at $M$ point which is also addressed by Xu \textit{et al.} \cite{Xu_2019}. Thus, this motivates to carry out ARPES study for finding more hole-like bands near E$_F$ at $M$ points. Yet, the hole-like bands near the Fermi energy at $\Gamma$ point and electron-like bands at $R$ points have already been reported in ARPES study \cite{Takane}. Thirdly, two band-crossing points are found at $\Gamma$ (bands 1, 2 \& 3) and $R$ (bands 1, 2, 3 \& 4) points in both figures. However, there is a shift in energy positions of band-crossing points found at $\Gamma$ and $R$ from $\sim$ 12 meV to -30 meV and from $\sim$ -194 meV to -186 meV, respectively, as observed while going from Fig. 2(a) - 2(b). These band-crossing points at DFT level seem to be three-fold degenerate at $\Gamma$ and four-fold degenerate at $R$ which are also similar to other theoretical and experimental works \cite{Tang,Gilbert,Chang,Takane}. The three-fold degeneracy has been associated with spin-1 fermion while the other four-fold degeneracy with double Weyl fermion in Ref.\cite{Tang,Gilbert,Chang,Takane}. Fourthly, in Fig. 2(b) as the band-crossing points $\Gamma$ and $R$ are lying in the range of -0.2$\leq\omega\leq$0.2 eV, this suggests that spin-1 and double Weyl fermionic QPs seem to have coherent weights (larger lifetime due to QP-QP interactions which is otherwise infinite in the case of non-interacting system). Consequently, at $\omega\sim$ -30 meV and $\omega\sim$ -186 meV the $\tau$ of QPs are evaluated as $\sim 10^{-9}$s  and $\sim 10^{-12}$s, respectively. However, $G_0W_0$ corrections have shown that the $Im \Sigma$ is almost zero for spin-1 fermionic QP at $\Gamma$ point indicating its $\tau$ to be almost infinite while $\tau$ of double Weyl fermionic QP at $R$ point is found to be $\sim 10^{-12}$s. Furthermore, the effective masses of QPs ($m^*$ values) as calculated from DFT+DMFT and $G_0W_0$ methods are given in Table I. Following the table almost equal values of $m^*$ seem to be possessed by all the three components of Co 3\textit{d} orbital at both $\omega$'s. Indicating that the QPs have become heavier due to electronic correlations, and further suggesting spectral weight transfers between coherent and incoherent states of these components in equal amount. This behavior can also be witnessed from Fig. 3(a) where all the three components of Co 3\textit{d} orbitals have negligible value of $Im \Sigma (\omega)$ around the E$_F$. Lastly, the $m^*$ values as calculated from DFT+DMFT method at both $\omega$ are containing the informations of both spin-1 and double Weyl fermionic QPs with QPs available at other $k$ points due to the fact that $Im \Sigma$ depends upon only $\omega$. Although, this is not the case with $m^*$ values as evaluated from $G_0W_0$ method due to the fact that here $Im \Sigma$ is a function of both $k$ \& $\omega$. Thus, 1.60 and 1.64 are the effective masses of spin-1 and double Weyl fermionic QPs found at $\Gamma$ and $R$ points, respectively.

Next, in order to see the effect of SOC on the spectral function in the presence of electronic correlations Fig. 4(a) - 4(b) are plotted. In Fig. 4(a) the DFT obtained bulk band structure and in Fig 4(b) momentum-resolved many-body spectral function calculated within DFT+DMFT at T = 100 K are shown. Both of them are plotted along high-symmetric $k$-directions for the energy window -0.5$\leq\omega\leq$0.2 eV with SOC inclusion. Here 8 bands are numbered in Fig. 4(a) due to the splitting of those 4 bands as found in Fig. 2(a), for instance, band 1 has split into 1 \& 1' bands. At first, it is found that the extra hole pockets at $M$ point got enhanced and the energy positions of the bands at $M$ has shifted from $\sim$ -16 meV to 23 meV in Fig. 4(b). Different fermions at both k-points are observed at DFT level  which is also reported in other works \cite{Tang,Sanchez,Takane,Pshenay}. It is said that in space group 198 due to the absence of inversion symmetry, SOC inclusion lifts the double degeneracy at non time-reversal (TR) invariant $k$-points while at TR invariant $k$-points the double degeneracy stays protected \cite{Cano}. Likewise, at $\Gamma$-point the six-fold degeneracy has split into two crossing points one with two-fold (created by 1 \& 1' bands) and other with four-fold degeneracy (created by bands 2, 2', 3 \& 3') in Fig. 4(a) \& 4(b). The two-fold degeneracy is associated with spin-1/2 Weyl fermion while four-fold degeneracy is associated with spin-3/2 RSW fermion. On moving from Fig. 4(a) to 4(b), change in energy positions of these band-crossing points at $\Gamma$ has been found. For instance, the four-fold degenerate point has moved from $\sim$ 30 meV to 4 meV in Fig. 4(b) while the two-fold  degenerate point has moved from $\sim$ -23 meV to -40 meV in Fig. 4(b). Similarly, at R-point in Fig. 4(a) \& 4(b), a crossing point with six-fold degeneracy (created by bands 2, 2', 3, 3', 4 \& 4') is found, which corresponds to double spin-1 excitations and its energy position has shifted from $\sim$ -186 meV to -182 meV. However, bands 1 \& 1' become degenerate just few energies below six-fold degenerate point and this too has shifted from $\sim$ -220 meV to -207 meV in Fig. 4(b). Moreover, in Fig. 4(b) the incoherent features seem to be enhanced at both band-crossing points where $\omega <$ -0.2 eV, and this can be validated from Fig. 3(b) where Im$\Sigma(\omega)$ is negligibly small. Existence of incoherent features at both $\Gamma$ and $R$ points suggesting the interactions between the QPs; affecting their lifetime. Thus, at $\omega \sim$ 4 and -40 meV, their $\tau \sim 10^{-9}$s while at $\omega \sim$ -180 and -207 meV, their $\tau \sim 10^{-12}$s. It is already known that an exact informations of newly fermionic QPs interactions at specific $k$-points cannot be retrieved from DFT+DMFT calculations. For which $G_0W_0$ corrections calculated the $\tau$ of newly fermionic QPs available at $\Gamma$ and $R$ points as $\sim 10^{-9}$s and $\sim 10^{-11}$s, respectively. Here again, the $m^*$ values as calculated from DFT+DMFT and $G_0W_0$ methods are given in Table II. From the table, it can be seen that DFT+DMFT has show almost same effective masses of the QPs $\sim$ 1.20. However, $G_0W_0$ has shown effective masses of spin-1/2 Weyl fermionic and spin-3/2 RSW fermionic QPs at $\Gamma$ to be same $\sim$ 1.6 whereas double spin-1 fermionic QPs found at $R$ point to have effective mass of $\sim$ 1.67.

Further for better visualizations, the enlarged pictures corresponding to the band-crossings found at $\Gamma$ and $R$ points in BZ, are discussed in the following figures for with and without SOC which are obtained from DFT and DFT+DMFT methods, respectively. Fig. 5 illustrates bulk band structure at $\Gamma$ point obtained from DFT both for (a) without SOC and (b) with SOC considerations in the energy window -0.04 $\leq \omega \leq$ 0.09 eV. On observing the Fig. 5(a), two linear bands (1 \& 3) and one flat band (2) which have a three-fold degenerate point located at $\sim$ 12 meV are found. On other hand in Fig. 5(b), due to SOC effect the bands 1, 2 \& 3 have split into three more bands 1', 2' \& 3' and created two band-crossing points. One of them is four-fold degenerate located at $\sim$ 30 meV which is created by 2, 2', 3 \& 3' bands where all these bands are almost linear at the point. The other one is two-fold degenerate located at $\sim$ -23 meV which is created by 1 \& 1' bands. These results are similar to other theoretical works \cite{Tang,Pshenay,Takane,Rao}. Then, two more linear band-crossing points (the marked ones) appear to be created by 1 \& 2' bands at $\sim$ -6 meV in $\Gamma - M$ direction while another at $\sim$ -4 meV in $\Gamma - R $ direction, respectively. However, other flat band-crossing point has been created (2 \& 2') at $\sim$ 16 meV in $\Gamma - R$ direction. These two band-crossing points in $\Gamma - R$ direction are recognized as type-I Weyl fermion by Tang \textit{et al.}\cite{Tang}. Moreover, due to the shift in the band crossing point at $\Gamma$ from $\sim$ 12 meV to 30 meV, the hole-type band (2 band) has also shifted in Fig. 5(b). This has increased the probability of states which will behave as hole-type. Next, we are showing the differences occurred in the band crossing point at $\Gamma$ after inclusion of SOC at DFT+DMFT level. For this Fig. 6(a) - 6(b) are plotted where momentum-resolved many-body spectral functions calculated within DFT+DMFT at T=100 K for with and without SOC effect are shown. Here in Fig. 6(a), the three-fold degenerate point is located at $\sim$ -30 meV and the hole pocket is found just in $\Gamma - R$ direction but not in $\Gamma - M$ direction which is unlike DFT result. Moreover, the two linear bands seem to possess coherent weight only while the flat band seems to possess both coherent and incoherent weights. Now moving to Fig. 6(b), as two band crossing points have generated at $\Gamma$ which are located at $\omega$ $\sim$ 4 meV and -40 meV corresponding to four-fold degenerate point and two-fold degenerate point, respectively. Interestingly, here the inclusion of SOC and Coulomb interactions have changed the curvatures of flat bands with enhanced incoherent weights. This has resulted in hole-pockets in both the directions $\Gamma - M$ and $\Gamma - R$. Furthermore, other band-crossing points can be clearly seen in Fig. 6(b) though their energy positions has shifted. Likewise, the two linear band-crossing points are located at $\sim$ -17.4 meV in $\Gamma - M$ direction and at $\sim$ -19.3 meV in $\Gamma - R$ direction. The flat band crossing is found to be shifted to $\sim$ 5 meV in $\Gamma - R$ direction.

Here, the differences occurred in band-crossing points at $R$ both from DFT and DFT+DMFT levels for with and without SOC effect are discussed. Accordingly, Fig. 7(a) - 7(b) and Fig. 8(a) - 8(b) are plotted for DFT and for DFT+DMFT at T=100 K, respectively. In Fig. 7(a), four linear bands (1, 2, 3 \& 4) are having a four-fold degenerate point which is located at $\sim$ -194 meV. On the contrary, SOC effect has split four linear bands into six linear bands (2, 2', 3, 3', 4 \& 4') whereas other 2 bands (1 \& 1') as parabolic are observed in Fig. 7(b). These six-fold degenerate point is located at $\sim$ -186 meV while two-fold degenerate parabolic bands are located at $\sim$ -220 meV. Here again in Fig. 7(b), two more linear band-crossing points are observed which are two-fold degenerate in $\Gamma-R$ direction where one is located at $\sim$ -205 meV and other is at $\sim$ -262 meV, respectively. These two band-crossing points are identified as type-II Weyl fermion by Tang \textit{et al.} \cite{Tang}. Next moving into Fig. 8(a), the four-fold degenerate point here is located at $\sim$ -186 meV. and the features are highly smeared suggesting the existence of incoherent weights; indicating larger QP-QP interactions. Then in Fig. 8(b), the six-fold degenerate point is located at $\sim$ -182 meV and 2-fold parabolic band-crossing point is at $\sim$ -207 meV. Moreover, the other two-fold linear band-crossings in $\Gamma - R$ direction are now located at $\sim$ -204 meV and $\sim$ -250 meV where features are again highly smeared.

In this part of discussion and results, the effects of slab thickness and electronic correlations over the surface states (SS) when projected to the (001) surface of CoSi along high symmetry lines $\overline{M}-\overline{\Gamma}-\overline{X}$ are discussed. Generally, it is said that hole and electron pockets in the bulk are possessed by nontrivial topology and due to which SS can be observed on the side surface \cite{Tang}. Based on this, the SS are expected to be observed on the side surface (001) of CoSi due to presence of electron and hole pockets in the bulk at $\Gamma$ and $R$ points in figures Fig. 2(a) - 2(b) and Fig. 4(a) - 4(b). Accordingly, Fig. 9(a) - 9(b) have been plotted, showing the surface band structures as calculated from DFT for (001) surface without SOC consideration when slab thicknesses are (a) 33.54 Bohr and (b) 100.64 Bohr, respectively. Two surface bands (I \& II) have been found at the projections of bulk states at $\Gamma$ and R points in Fig. 9(a) - 9(b), respectively.,  marked in solid lines (red color). This is similar to the electronic spectra for (001) surface as reported by Tang \textit{et al.} \cite{Tang}. For explaining the changes occurred in the SS while going from Fig. 9(a) - 9(b) few elliptical markings have been numbered in the plots. In Fig. 9(a), it is observed that at marking 1 surface bands I and II both are crossing the Fermi level (E$_F$) separately while at marking 2 only surface band II has crossed the E$_F$. Next, at marking 3 both surface bands I \& II merged together and cross the E$_F$ as a single band whereas in marking 4 none of them crosses the E$_F$ and remain below the E$_F$. Lastly, at marking 5 only surface band I crosses the $\Gamma$ point at $\sim$ 166 meV. Moving to Fig. 9(b), it is found that at marking 1 both surface bands merged together crosses the E$_F$ as single band while at marking 2 this time the surface band I slightly shifts above the E$_F$. Then, at marking 3 both the surface bands crosses the E$_F$ separately. At marking 4 surface band II cross over the E$_F$ while surface band I remains below the E$_F$. This time at marking 5 and marking 6 both surface bands I \& II crosses the $\Gamma$ at $\sim$ 47 meV and 130 meV, respectively. All these points suggest that there are no major changes on the number of crossings at E$_F$ of the surface bands I and II. Based on this, it can be concluded that SS are not much affected with the change in slab thickness in the energy window of -50$\leq \omega \leq$50 meV. Furthermore, the momentum-resolved many-body spectral functions for (001) surface are also plotted for with and without SOC effect at T = 100 K in Fig. 10(a) - 10(b), respectively. Due to high computational cost, for the calculation of surface states from DFT+DMFT method the slab thickness with 33.54 Bohr is only chosen. On looking at Fig. 10(a) - 10(b), one can find that here again the surface states seem to emerge from the projections of bulk states at $\Gamma$ and $R$ points as expected. In Fig. 10(a) \& 10(b), both the surface bands appear to cross the E$_F$ at marking 4 unlike DFT results. Further, on moving from Fig. 10(a) to 10(b), the incoherency of the spectrum (smeared features) appears to enhance specifically in the region -0.1$\leq\omega\leq$-0.2 eV. This suggests that these surface states are interacting and due to which their lifetime is also affected. For instance, from Fig. 10(a) and Fig. 10(b) at $\omega$ = 0  the $\tau \sim 10^{-8}$s.

\section{Conclusion}

In this work, the electronic spectra of bulk and (001) surface of CoSi have been studied by using advanced DFT+DMFT method at T = 100 K for both with and without SOC inclusions, respectively. From the DFT+DMFT calculations for the bulk states one extra hole pocket is found at $M$ points in BZ. All the newly discovered fermions are observed similar to other theoretical and experimental reports. The dispersion curves of Bulk CoSi have shown the both coherent and incoherent features. Further suggesting the QP-QP interactions which is resulting into affecting their lifetime. For example, $\tau$ for QPs at $\omega \sim $ -30 and -186 mev are found to be $\sim 10^{-9}$ s and $\sim 10^{-12}$ s, respectively when SOC is not considered. However, $G_0W_0$ has given $\tau$ for spin-1 fermionic QP at $\Gamma$ as infinite while for double Weyl fermionic QP at $R$ point as $\sim 10^{-12}$ s. Their effective masses are also calculated by using $G_0W_0$ method as $\sim$ 1.60 and 1.64. Similarly, the $\tau$ and $m^*$ values of other fermionic QPs are also calculated by using both DFT+DMFT and $G_0W_0$ method under SOC inclusion. Furthermore, at T = 100 K the electronic spectra of (001) surface have also shown both coherent and incoherent features. Consequently, for both noc SOC and SOC inclusions at $\omega$ = 0, $\tau$ $\sim 10^{-8}$s has been calculated.

\begin{figure*}
\subfloat[\label{ 1}]{%
  \includegraphics[width=0.3\linewidth]{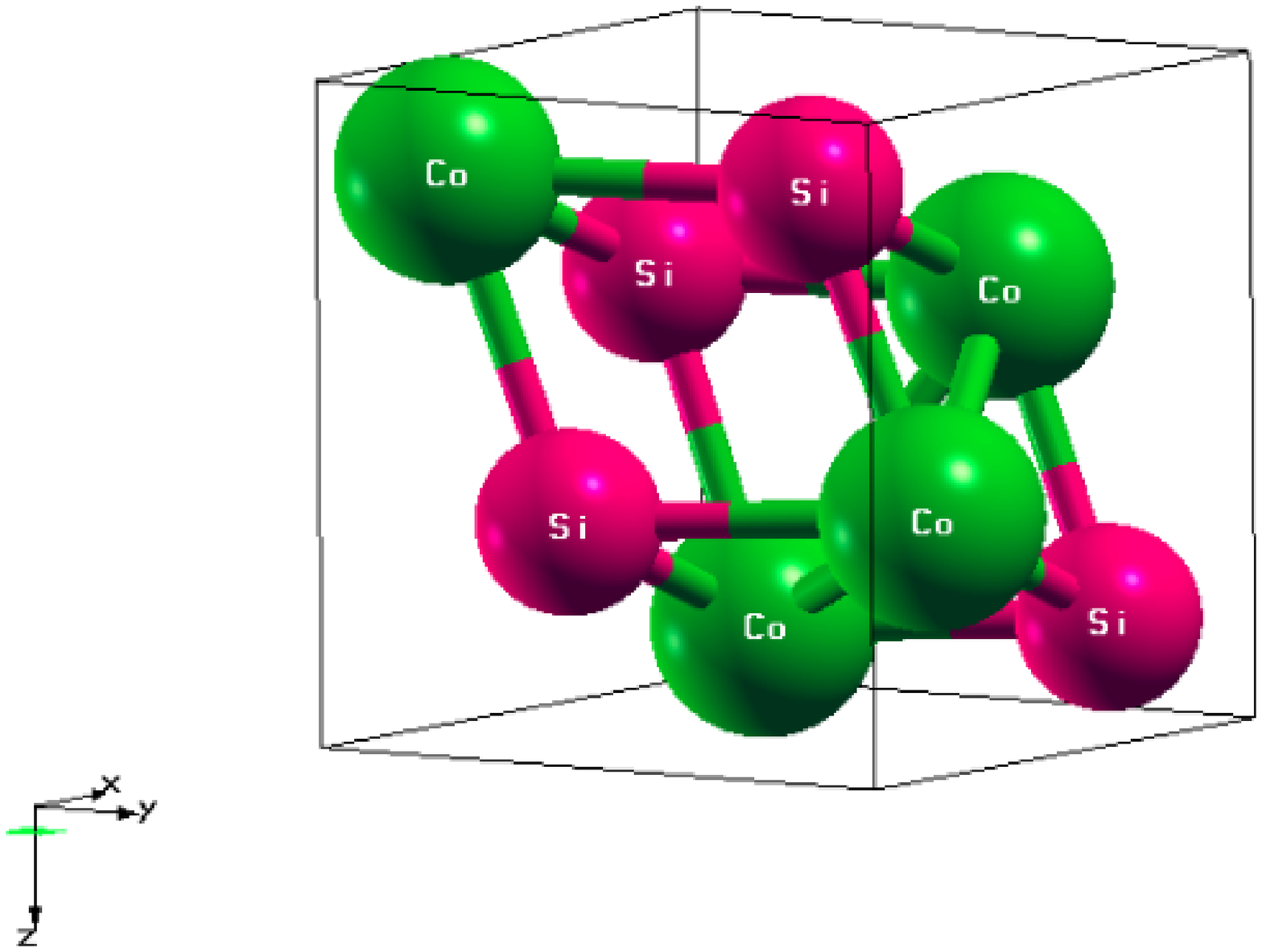}%
}\hfill  
\subfloat[\label{ 2}]{%
  \includegraphics[width=0.3\linewidth]{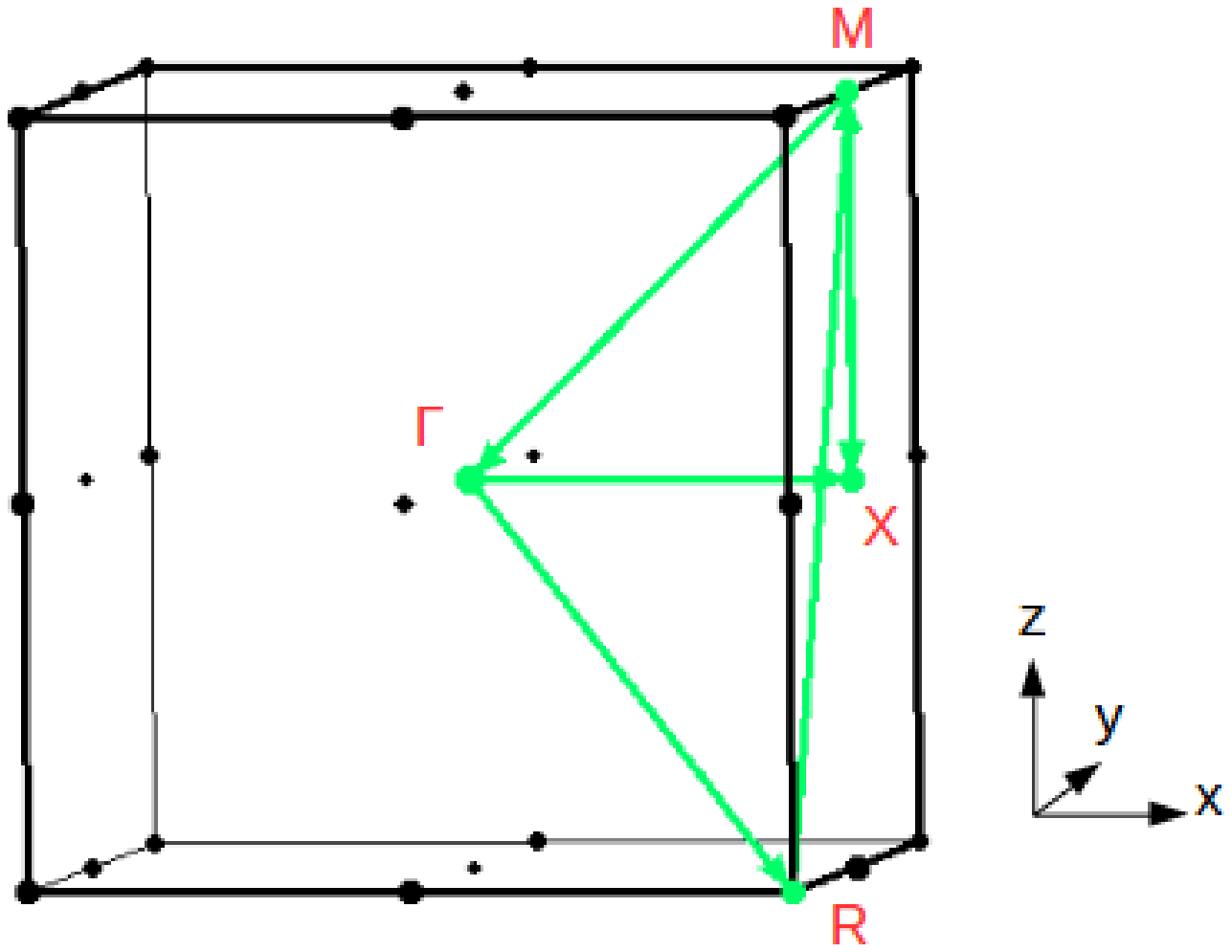}%
}\hfill
\subfloat[\label{ 3}]{%
  \includegraphics[width=0.3\linewidth]{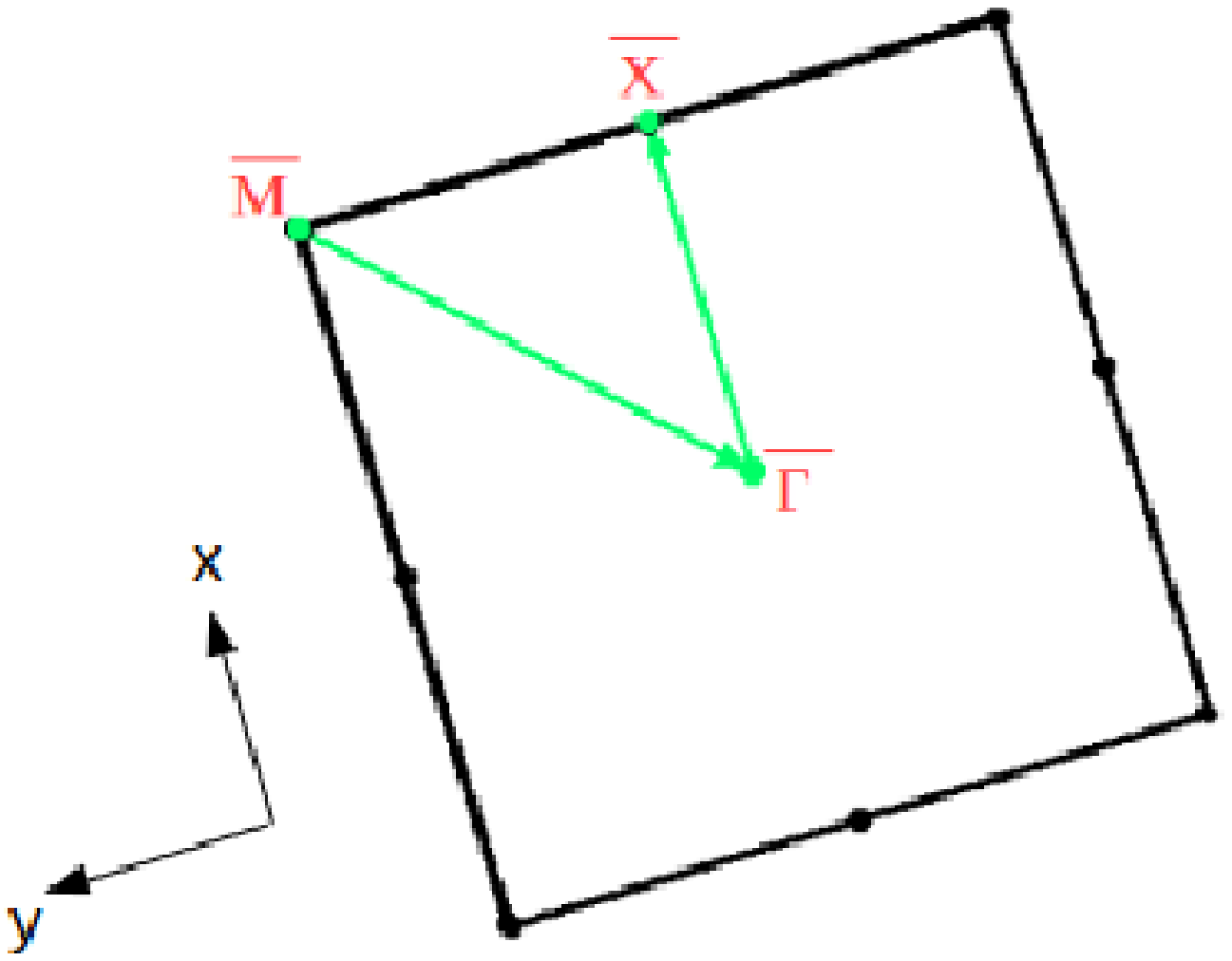}%
}\hfill
\caption{\small{(colour online) (a) The crystal structure of CoSi where bigger sphere is denoting Co atom while smaller one is denoting Si atom, (b) 3D-Brillouin zone (BZ) for CoSi showing some high-symmetric $k$-directions, (c) the projected BZ of 001 surface of CoSi. }}
\label{4}
\end{figure*}

\begin{figure*}
\subfloat[\label{ 5}]{%
  \includegraphics[height=1.4in, width=2.8in]{fig2_1.eps}%
}\hfill  
\subfloat[\label{ 6}]{%
  \includegraphics[height=1.75in,width=4in]{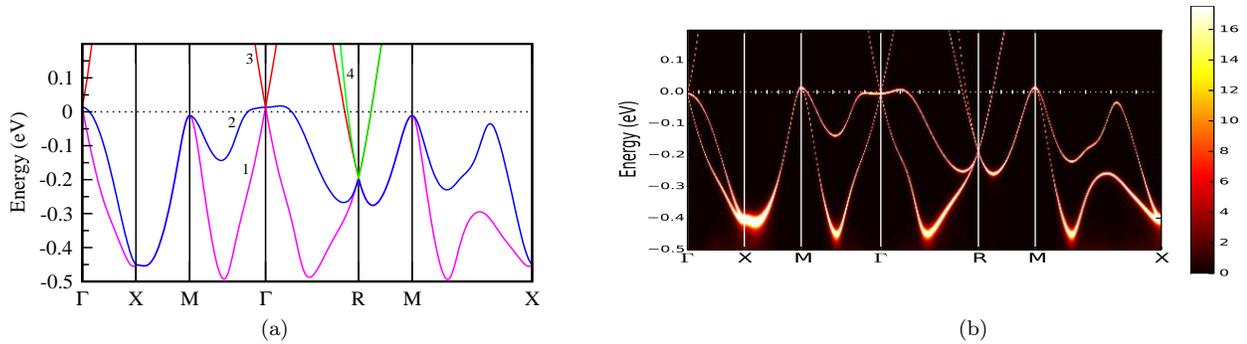}%
}\hfill
\caption{\small{(colour online) (a) Bulk band structure of CoSi without SOC as obtained from DFT calculation, and (b) Momentum-resolved many-body spectral function without SOC at T = 100 K obtained from DFT+DMFT. Zero energy represents the Fermi level. }}
\label{7}
\end{figure*}

\begin{table*}
%\captionsetup{font=footnotesize}
\caption{\footnotesize{Calculated $m^*$ for different $\omega $ corresponding to $\Gamma$ and $R$ points without SOC consideration at T = 100 K as evaluated from DFT+DMFT method for three orbital components of Co 3\textit{d} orbital and other at T=0 K as evaluated from $G_0W_0$ method, respectively. Different $\omega$ values are corresponding to energy positions of band-crossing points as obtained from Fig. 2(b) }}
\label{tab.1}
\begin{center}
\setlength{\tabcolsep}{10pt}
\footnotesize
\begin{tabular}{lcccccr}
\hline
\hline
 &  & Orbital component (DFT+DMFT) &  &  & & $G_0W_0$ \\
 
   & $z^2$ & x$^2$-y$^2$/xy  & xz/yz &  & \\
 \cline{2-5}
\\
 $k$-points ($\omega$(meV)) &  $m^*$  & $m^*$ & $m^*$ &  &  & $m^*$\\
  
\hline 
 
 $\Gamma$ ($\omega\sim$ -30) & 1.16 & 1.19 & 1.20 & &  &1.60\\
$R$ ($\omega\sim$ -186) & 1.16 & 1.20 & 1.23 &  &  &1.64  \\
%-406.0 & 1.18, 0.84 & 1.24, 0.81 & 1.26, 0.80 \\

\hline
\hline
\end{tabular}
\end{center}
\end{table*}

\begin{figure*}
  \begin{center}
   \includegraphics[width=2.3in]{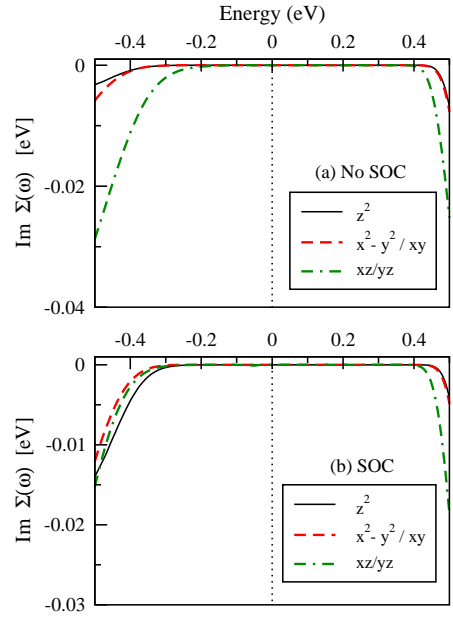}
  \end{center}

\caption{\small{(colour online) Imaginary part of self-energy ($Im \Sigma(\omega)$) as a funtion of energy at T = 100 K for three components of 3\textit{d} orbitals of Co (a) without SOC and (b) with SOC. Zero energy corresponds to the Fermi level (dotted line). }}
 
\end{figure*}

\begin{figure*}
\subfloat[\label{ 8}]{%
  \includegraphics[height=1.5in, width=2.7in]{fig4_1.eps}%
}\hfill  
\subfloat[\label{ 9}]{%
  \includegraphics[height=1.9in,width=4in]{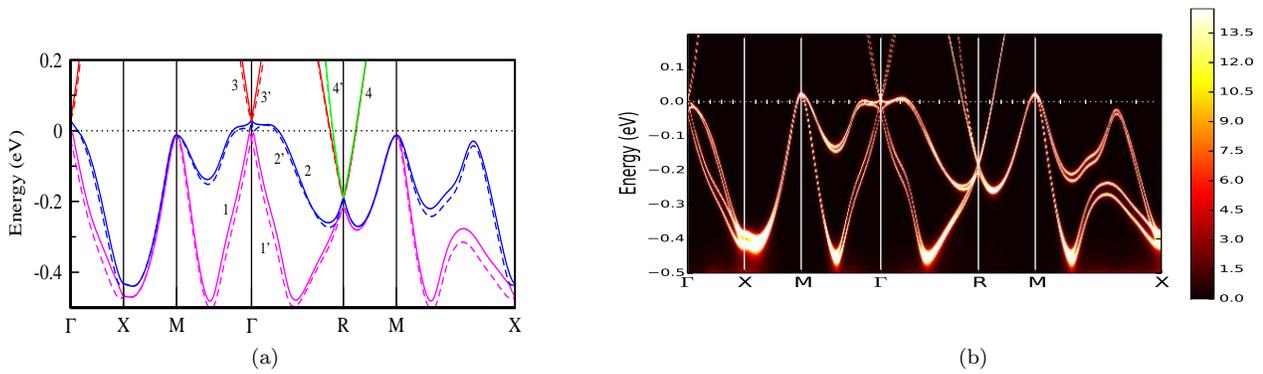}%
}\hfill
\caption{\small{(colour online) (a) Bulk band structure of CoSi with SOC consideration as obtained from DFT calculation, and (b) Momentum-resolved many-body spectral function with spin-orbit coupling consideration at T = 100 K obtained from DFT+DMFT. Zero energy represents the Fermi level. }}
\label{10}
\end{figure*}

\begin{table*}
%\captionsetup{font=footnotesize}
\caption{\footnotesize{Calculated $m^*$ for $\Gamma$ and $R$ points with SOC consideration at T = 100 K as evaluated from DFT+DMFT method for three orbital components of Co 3\textit{d} orbital and other at T=0 K as evaluated from $G_0W_0$ method, respectively. Different $\omega$ values are corresponding to energy positions of band-crossing points as obtained from Fig. 4(b). }}
\label{tab.2}
\begin{center}
\setlength{\tabcolsep}{10pt}
\footnotesize
\begin{tabular}{lccccr}
\hline
\hline
 &  & Orbital component (DFT+DMFT) &  &  & $G_0W_0$ \\
 
   & $z^2$ & x$^2$-y$^2$/xy  & xz/yz &  & \\
 \cline{2-5}
\\
 $k$-point ($\omega$(meV)) &  $m^*$  & $m^*$ & $m^*$ &  &  $m^*$\\
  
\hline 
 
  $\Gamma$ ($\omega\sim$ 4) & 1.18 & 1.20 & 1.20 & & 1.60   \\
 $\Gamma$ ($\omega\sim$ -40) & 1.18 & 1.19 & 1.20 & & 1.61\\
$R$ ($\omega\sim$ -180) & 1.18 &  1.20 & 1.20 & & 1.67 \\
$R$ ($\omega\sim$ -207) & 1.19 & 1.20 & 1.20 & & 1.64\\
\hline
\hline
\end{tabular}
\end{center}
\end{table*}

\begin{figure*}
  \begin{center}
   \includegraphics[width=2in]{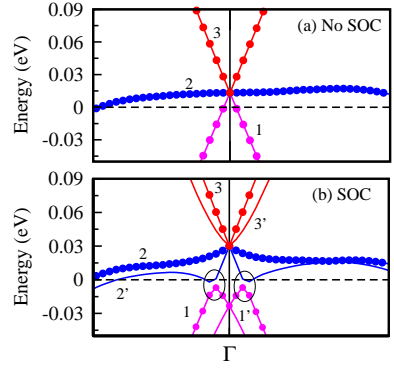}
  \end{center}

  \caption{\small{(colour online) Bulk band structure of CoSi at $\Gamma$ point obtained from DFT both for (a) without SOC and (b) with SOC. Zero energy corresponds to the Fermi level (dashed line). Two elliptical markings are drawn for the explanation purpose.}}
 
\end{figure*}

\begin{figure*}
\subfloat[\label{ 10}]{%
  \includegraphics[height=1.8in, width=2.8in]{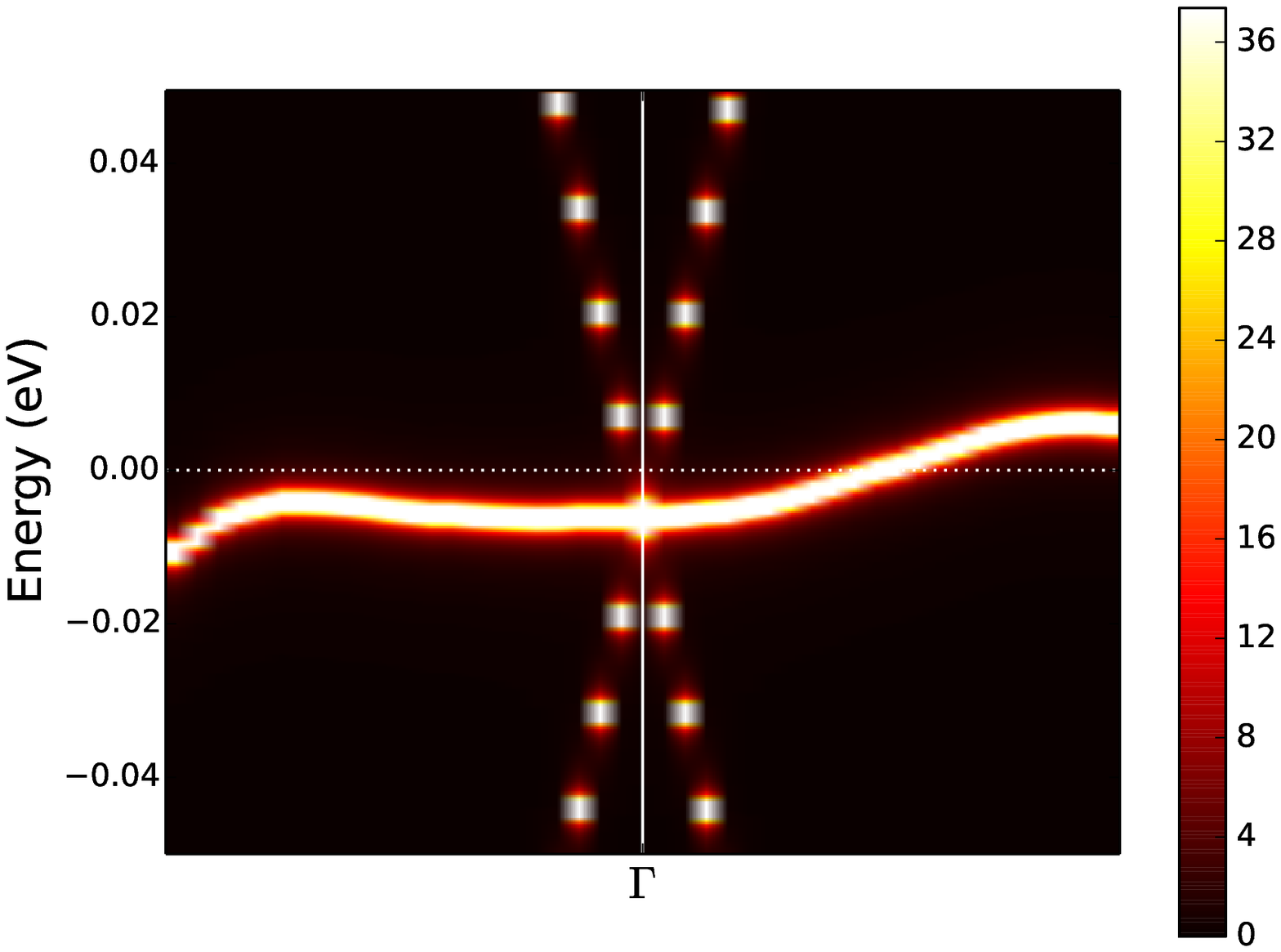}%
}\hfill  
\subfloat[\label{ 11}]{%
  \includegraphics[height=1.8in, width=2.8in]{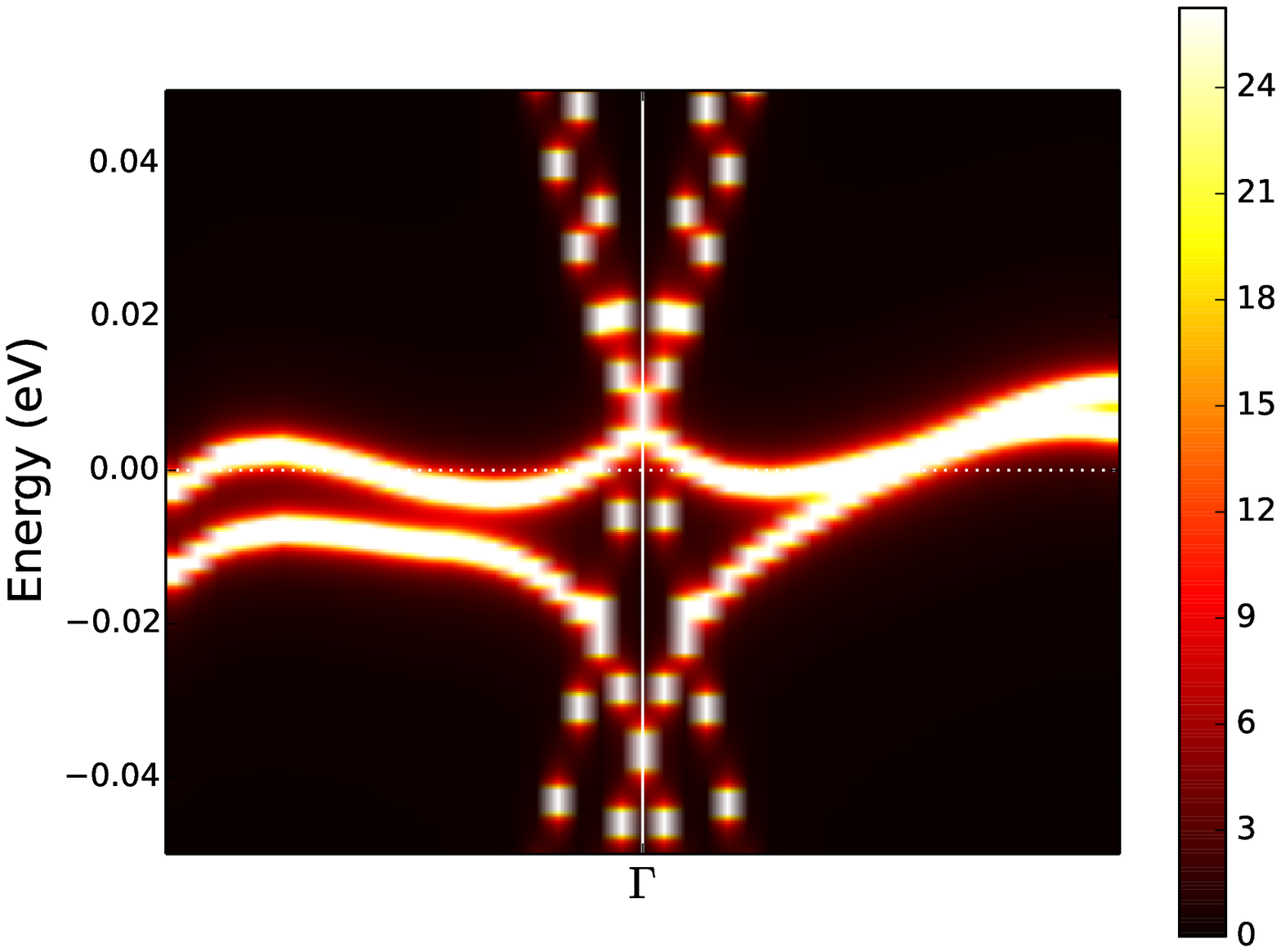}%
}\hfill
\caption{\small{(colour online) Momentum-resolved many-body spectral function at $\Gamma$ point for the T = 100 K obtained from DFT+DMFT both for (a) without SOC and (b) with SOC. Zero energy represents the Fermi level. }}
\label{12}
\end{figure*}

\begin{figure*}
  \begin{center}
    \includegraphics[width=2in]{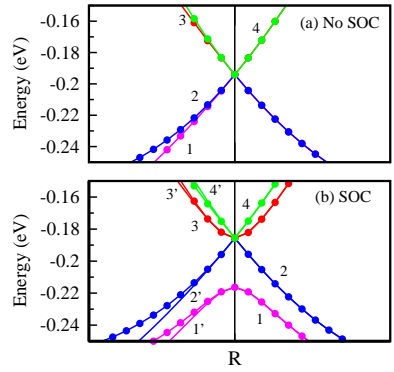}
  \end{center}

  \caption{\small{(colour online) Bulk band structure of CoSi at $R$ point obtained from DFT both for (a) without SOC and (b) with SOC. }}
 
\end{figure*}
\begin{figure*}
\subfloat[\label{ 13}]{%
  \includegraphics[height=1.8in, width=2.8in]{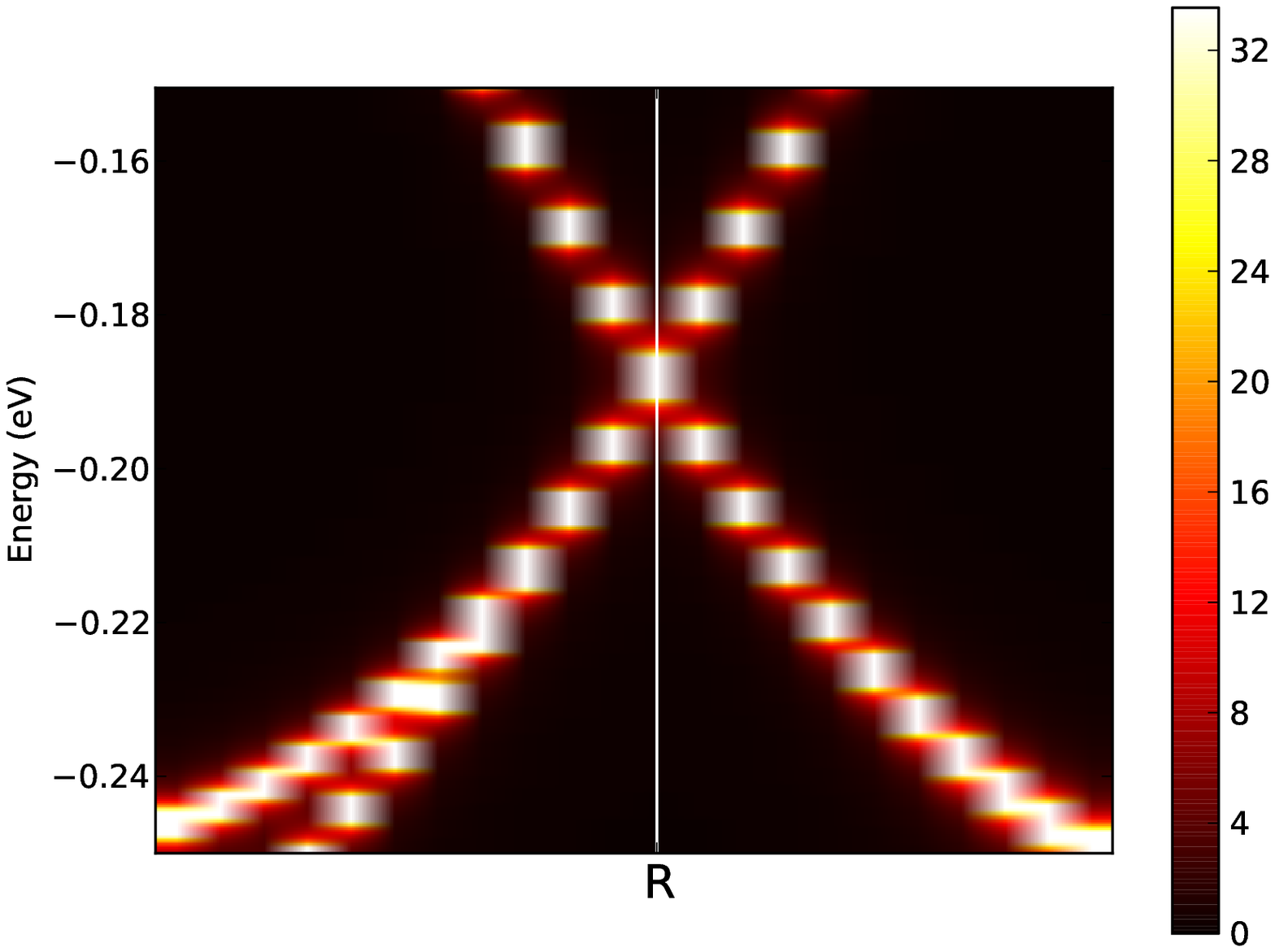}%
}\hfill  
\subfloat[\label{ 14}]{%
  \includegraphics[height=1.8in, width=2.8in]{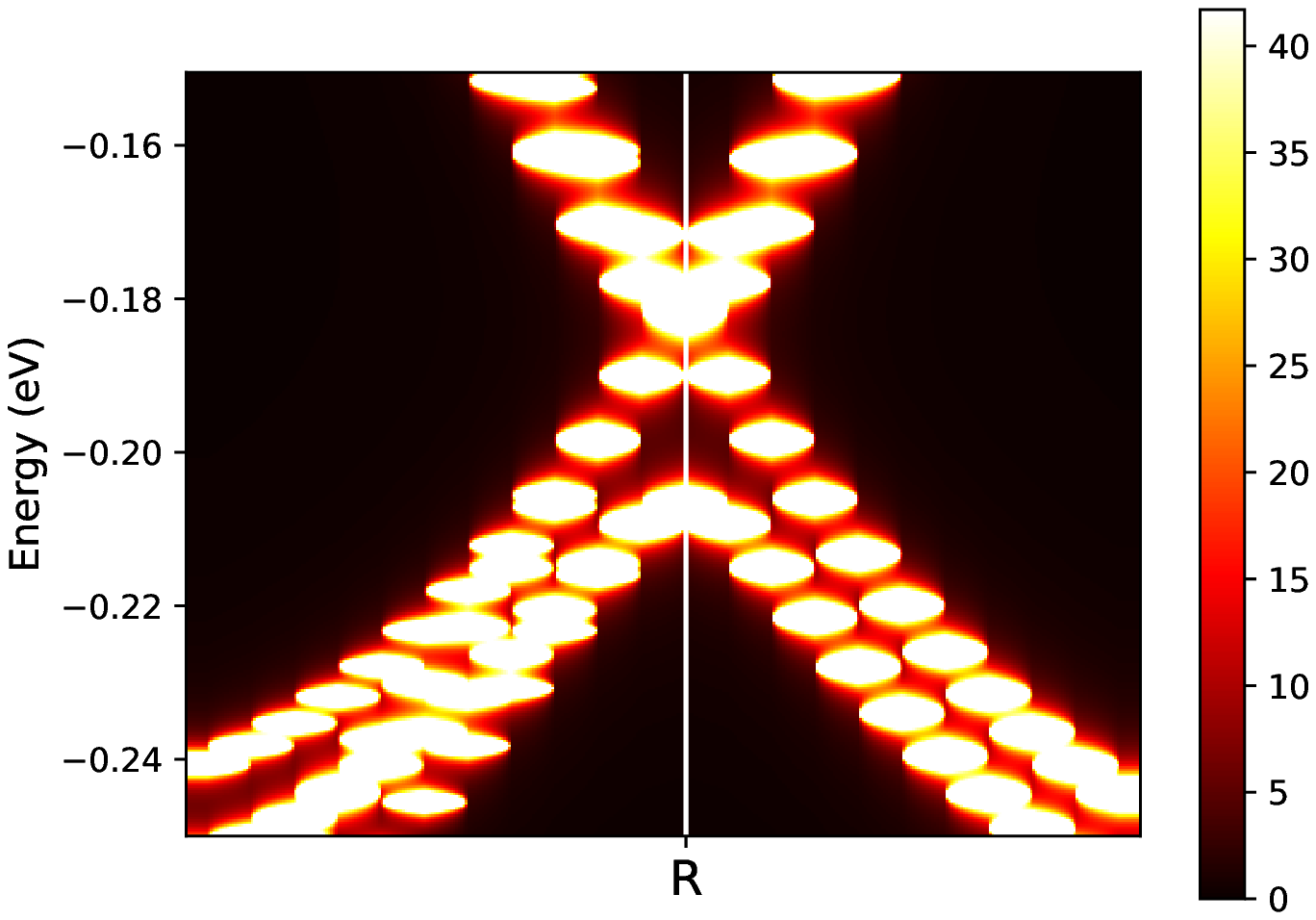}%
}\hfill
\caption{\small{(colour online) Momentum-resolved many-body spectral function at T = 100 K obtained from DFT+DMFT both at $R$ point for (a) without SOC and (b) with SOC. }}
\label{15}
\end{figure*}

\begin{figure*}
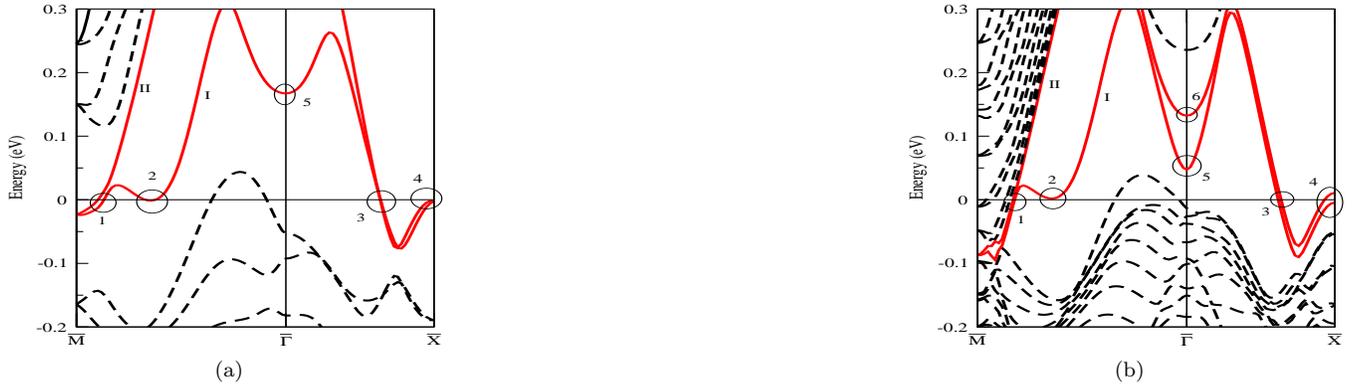

\subfloat[\label{16}]{%
\includegraphics[height=2in, width=2.3in]{fig9_1.eps}
}\hfill
\subfloat[\label{17}]{%
\includegraphics[height=2in, width=2.3in]{fig9_2.eps}
}  
\caption{\small{(colour online) Band structure of (001) surface of CoSi obtained from DFT both for without SOC inclusion when the slab thickness is (a) 33.54 bohr and (b) 100.64 bohr. The surface states are marked with solid lines (red color) and bulk states are marked with dashed lines (black color). Zero energy represents the Fermi level. Elliptical markings are shown for explanation purpose. }}
 \end{figure*} 
 
\begin{figure*}
\subfloat[\label{18}]{%
 \includegraphics[height=2in, width=2.8in]{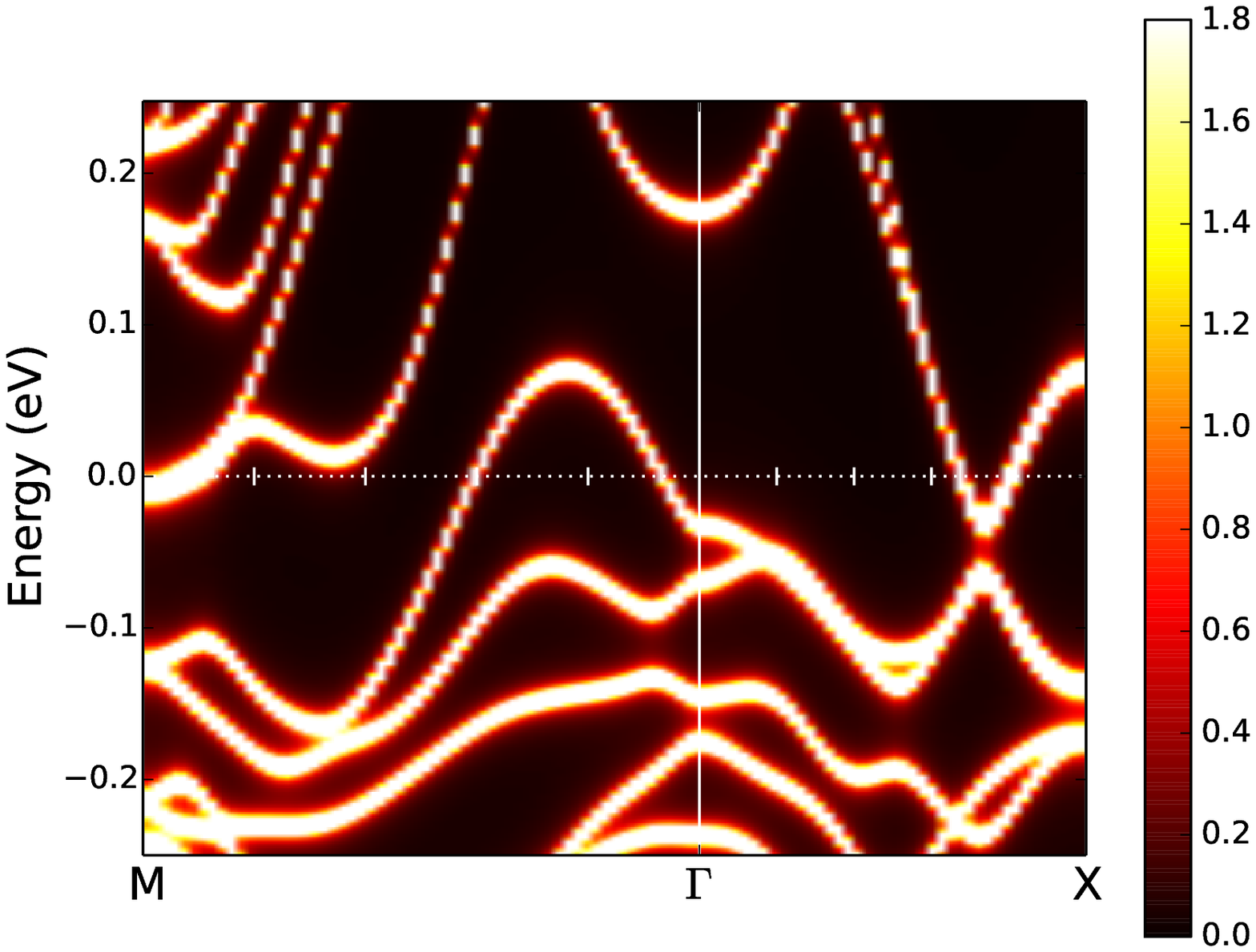}
}\hfill
\subfloat[\label{19}]{% 
\includegraphics[height=2in, width=2.8in]{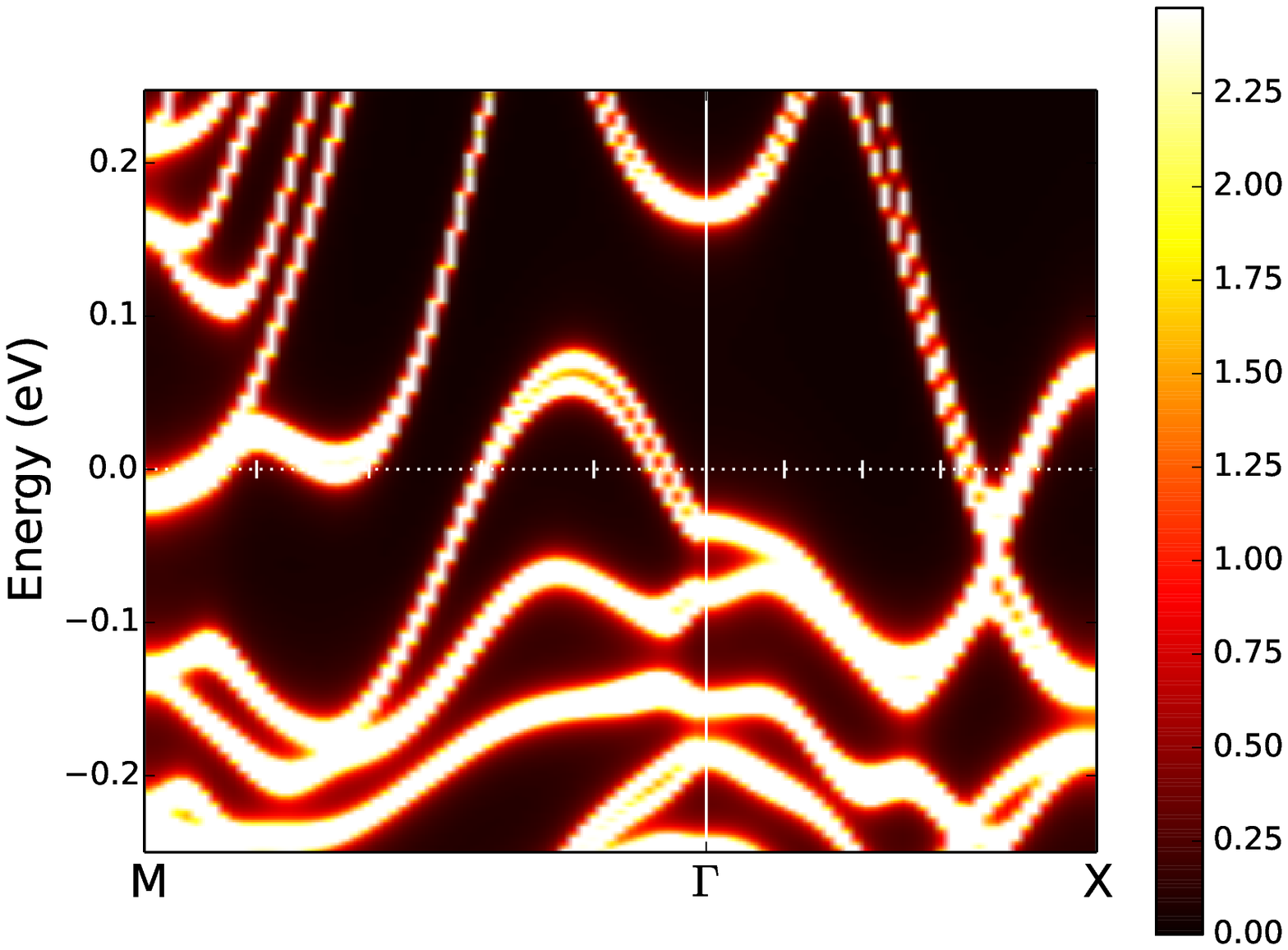}
}\caption{\small{(colour online) Momentum-resolved many-body spectral function for (001) surface of CoSi at T = 100 K obtained from DFT+DMFT both for (a) without SOC and (b) with SOC. Zero energy represents the Fermi level. }}
 \end{figure*}

\end{document}